\documentclass[a4paper,twocolumn,11pt,floatfix,accepted=2026-07-29]{quantumarticle}

\pdfoutput=1

\usepackage[utf8]{inputenc}
\usepackage[english]{babel}
\usepackage[T1]{fontenc}
\usepackage{hyperref}
\usepackage{tikz}
\usepackage{lipsum}
\usepackage[numbers,sort&compress]{natbib}
\usepackage{amssymb}
\usepackage{xcolor}
\usepackage{natbib}
\usepackage{graphicx}
\usepackage{color}
\usepackage{braket}
\usepackage{amsmath}
\usepackage{array}
\usepackage{tabularx}
\usepackage{mathtools}
\usepackage{commath}
\usepackage{bbold}
\usepackage{svg}
\usepackage{soul}

\begin{document}

\title{Robust topological quantum state transfer with long-range interactions in Rydberg arrays}

\author{Siri Raupach}
\author{Beatriz Olmos}
\author{Mathias B. M. Svendsen}
\affiliation{Institut f\"ur Theoretische Physik, Universit\"at Tübingen, Auf der Morgenstelle 14, 72076 T\"ubingen, Germany}

\begin{abstract}

We develop a theoretical framework for fast, robust and high-fidelity topological quantum state transfer in one-dimensional systems with long-range couplings, motivated by chains of Rydberg atoms with dipole–dipole interactions. Such long-range interactions naturally give rise to extended Su–Schrieffer–Heeger and Rice–Mele models supporting topologically protected edge states. We show that these edge states enable high-fidelity edge-to-edge excitation transfer using both time-independent protocols, based on coherent edge state dynamics, and time-dependent protocols, based on adiabatic modulation of system parameters. Long-range couplings play a central role by enhancing the relevant energy gaps, leading to a substantial improvement in transfer efficiency compared to nearest neighbour models. The resulting transfer is robust against positional disorder, reflecting its topological origin and highlighting the potential of long-range interacting platforms for reliable quantum state transfer.
\end{abstract}

\maketitle

\section{Introduction}

A key ingredient in efficient quantum information processing is the reliable transfer of quantum states between distant nodes of a quantum network \cite{divincenzo2000,kimble2008}. Numerous quantum state transfer (QST) protocols have been proposed across a wide range of platforms, including spin chains \cite{Huang2018,Korzekwa2014,Bose2003,Christandl2004}, superconducting qubit networks \cite{Karamlou2022,Braumuller2022}, cavity arrays \cite{Almeida2016}, and optical systems \cite{Ru2021}. These protocols broadly fall into two classes: time-independent schemes based on free evolution under an engineered Hamiltonian, and time-dependent schemes relying on controlled parameter modulation. While time-dependent protocols often exploit adiabatic evolution, the long timescales required to satisfy adiabaticity render them susceptible to decoherence and disorder, limiting achievable transfer fidelities. 


An effective strategy to mitigate these limitations is to exploit symmetry-protected topological phases, which can host edge states that are robust against local perturbations and disorder \cite{Schnyder2008,Ryu2010}. Such edge states enable edge-to-edge QST across an entire system and have been explored in a variety of topological settings. In one dimension, topological QST has been proposed in the Su–Schrieffer–Heeger (SSH) \cite{lang2017,estarellas2017,Mei2018,Yuan2023,Huang2022,Palaiodimopoulos2021} and Rice–Mele models \cite{Qi2020,Xing2022}, using both time-dependent adiabatic protocols and time-independent approaches based on coherent edge state dynamics, such as Rabi flopping \cite{Bello2016,Longhi2019}. Related ideas have also been investigated in two-dimensional topological lattices \cite{Wei2023,Dlaska2017}.

Most existing studies of topological QST focus on nearest neighbour models. Recent works suggest that long-range couplings can both destabilize and facilitate topological phases \cite{Gong2016}, and may even give rise to emergent topological phenomena with fractional or higher valued topological invariants (beyond $\pm 1$) \cite{Viyuela2016,Kim2024,Li2019,Hsu2020} and novel types of corner states \cite{Li2020,Olekhno2022,Qin2025}. Whether and how such long-range interactions can be harnessed to improve the performance of topological QST, rather than degrade it, remains an open question. Rydberg atoms trapped in optical tweezers provide a natural theoretical platform to explore this regime owing to their strong and inherently long-range dipole–dipole interactions and long coherence times \cite{Schaus2012,Kaufman2021}. Moreover, the high degree of control over individual tweezer positions enables the exploration of a broad range of lattice geometries and interaction patterns \cite{Bluvstein2022,Browaeys2020}. Rydberg lattices have been employed to theoretically predict and experimentally observe edge states in both one- and two-dimensional topological models \cite{Weber2022,Weber2018,Deleseleuc2019,Li2021,Svendsen2025}, as well as to simulate robust topological pumping \cite{Svendsen2024,Trautmann2024}.

In this work, we investigate quantum state transfer in a topological one-dimensional chain in the experimentally relevant regime of Rydberg atom arrays, where interactions are inherently long-range and of dipolar form. Rather than treating these long-range couplings as perturbations to idealized nearest-neighbor SSH or Rice–Mele models, we incorporate them directly into a time-dependent state-transfer protocol and analyze their impact on both dynamics and performance. We show that long-range interactions qualitatively modify the transfer process and can, in fact, act as a resource: they enhance transfer speed and improve robustness against disorder in regimes relevant for current experiments. Our results therefore bridge the gap between idealized topological state-transfer protocols and realistic implementations in Rydberg platforms, and provide a systematic framework for understanding how intrinsic interaction profiles influence adiabatic topological transport. More broadly, this work highlights how long-range interactions can reshape established paradigms of topological quantum state transfer and suggests new avenues for optimizing such protocols in experimentally constrained quantum systems.


\section{Single-sided edge states in one dimension}

\begin{figure}[t]
    \centering
    \includegraphics[width =\columnwidth]{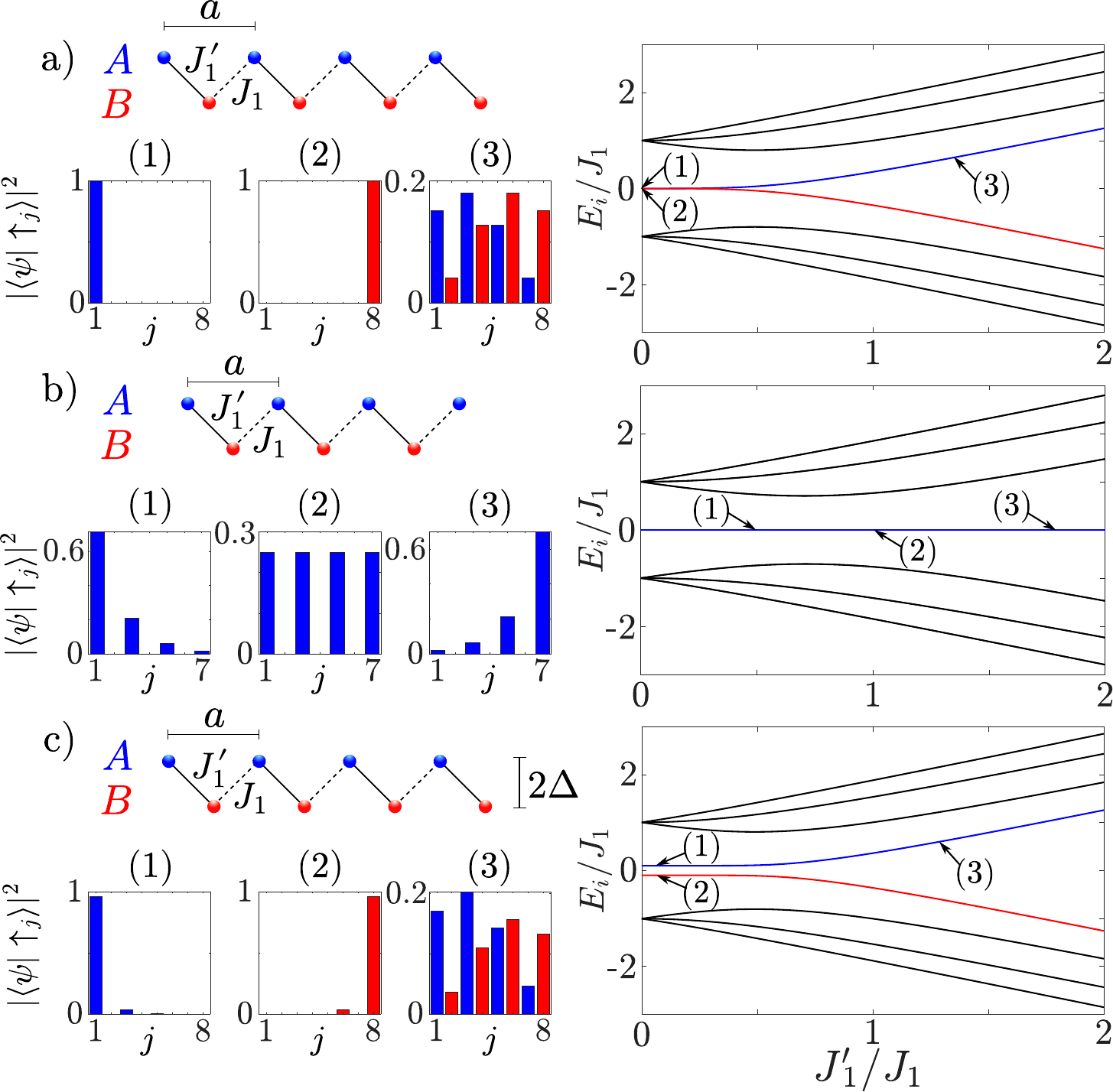}
    \caption{\textit{Single-sided topological edge states}. a) SSH model with even number of sites. The energy spectrum (right) depicted as a function of the ratio $J_1'/J_1$ shows that in the topologically trivial regime ($J_1'/J_1>1$) the mid-gap states occupy the bulk (3), while for the specific case $J_1'/J_1=0$, two single-sided edge states appear, (2) and (1). b) SSH model with odd number of sites. Here, single-sided edge states exist in both the topological (1) and trivial (3) regimes. At the transition point, where $J_1'=J_1$, the mid-gap state is an equal superposition of all sites on sublattice $A$ (2). c) Rice-Mele model. The mid-gap states are split due to the energy shift $\hbar\Delta$, and single-sided edge states (1) and (2) are found in the regime where $J_1'/J_1<1$, while again for $J_1'/J_1>1$ the mid-gap states occupy the bulk (3).}
    \label{TopologicalModels}
\end{figure}
The SSH model is a canonical model to study topological phases of matter in one dimension. It describes a single (spinless) fermionic particle hopping between neighbouring sites on a lattice of $N$ sites divided into two sublattices $A$ and $B$ [see Figure \ref{TopologicalModels}a)]. Each unit cell consists of one site of each sublattice, with neighbouring cells separated by a distance $a$. The particle can hop within its unit cell and to a neighbouring unit cell with coupling rates $J_1'$ and $J_1$, respectively. The Hamiltonian of the system is
\begin{equation}
    \hat{H}_\mathrm{SSH} = \hbar \!\!\sum_{n=1}^{\lfloor N/2\rfloor}\!\! J_1'\hat{a}_n^\dagger \hat{b}_n+\hbar\!\!\!\!\!\sum_{n=1}^{\lceil N/2\rceil -1}\!\!\!\!\! J_1\hat{b}_n^\dagger \hat{a}_{n+1} + \textrm{h.c.},
    \label{eq:HamiltonianSSH}
\end{equation}
where $\hat{a}_n^\dagger$ ($\hat{a}_n$) and $\hat{b}_n^\dagger$ ($\hat{b}_n$) are the fermionic creation (annihilation) operators for sublattices $A$ and $B$ in unit cell $n$, respectively.  

The topological properties of the SSH model can be inferred from its bulk Hamiltonian, obtained by imposing periodic boundary conditions
\begin{equation}
    \hat{H}_\mathrm{SSH} =\!\!\!\!\sum_{\substack{k\in \text{FBZ}\\ \alpha,\beta\in A,B}}\!\!\!\! \hat{c}^\dag_{\alpha}(k) h_{\alpha\beta}(k) \hat{c}_{\beta}(k).
    \label{eq:BlochHamiltonian}
\end{equation}
Here, the operators in equation \eqref{eq:HamiltonianSSH} have been Fourier transformed as
\begin{equation}
    \hat{c}_\alpha(k)=\frac{1}{\sqrt{\lfloor N/4\rfloor}}\sum_{n=1}^{\lfloor N/4\rfloor}\! e^{-\mathrm{i}kna}\,\hat{c}_\alpha(n),
\end{equation}
with $\hat{c}_A(n)\equiv \hat{a}_n$ and $\hat{c}_B(n)\equiv \hat{b}_n$. In momentum space, the Hamiltonian takes the irreducible form
\begin{equation}
     h(k) = \hbar\!
    \begin{pmatrix}
        0 && n(k) \\ 
        n^*(k) && 0
    \end{pmatrix}
    \!,
\end{equation}
with $n(k) = J_1'+J_1e^{ika}$.

For real coupling rates, $h(k)$ exhibits time reversal symmetry $\hat{\mathcal{T}}h(k)\hat{\mathcal{T}}^{-1}=h(-k)$, particle-hole symmetry $\hat{\mathcal{C}}h(k)\hat{\mathcal{C}}^{-1}=-h(-k)$ and chiral symmetry $\hat{\mathcal{S}}h(k)\hat{\mathcal{S}}^{-1}=-h(k)$, placing the model in the BDI symmetry class with a $\mathbb{Z}$-type topological invariant \cite{Ryu2010}. The relevant topological invariant is the winding number
\begin{equation}
    \nu = \frac{i}{2\pi}\int_{\mathrm{FBZ}}\mathrm{d}k \,n(k)\,\partial_k n^*(k),
\end{equation}
which takes the value $\nu=0$ for $|J_1'|>|J_1|$ (topologically trivial) and $\nu=1$ for $|J_1'|<|J_1|$ (topologically non-trivial).

The topologically non-trivial phase is characterized by eigenstates of the Hamiltonian in equation \eqref{eq:HamiltonianSSH} that are exponentially localized at the lattice boundaries, known as edge states. Their energies are well separated from the bulk spectrum approaching zero in the thermodynamic limit, $N\rightarrow \infty$. We refer to these states here as mid-gap states. They are robust against local perturbations that preserve the underlying symmetries of the Hamiltonian. As we described earlier, time-dependent QST exploits adiabatic evolution, connecting two instantaneous eigenstates of the Hamiltonian by slowly varying its parameters. Achieving edge-to-edge transfer in the SSH model requires edge eigenstates that are localized on a single boundary. In the following, we outline the conditions under which such single-sided edge eigenstates arise.

\subsection{Even SSH model, fine-tuned}
In an SSH chain with an even number of lattice sites, the topologically non-trivial phase hosts two mid-gap states localized near the opposite boundaries. To describe these edge states, one may adopt an exponentially localized ansatz for the single-particle wavefunction \cite{Mei2018,Asboth2016},
\begin{equation}
    \ket{\psi_E} =\mathcal{N}\sum_{n=1}^{N/2}\lambda^n\left(\alpha \hat{a}_n^\dagger+\beta \hat{b}_n^\dagger\right)\ket{G},
    \label{eq:ansatz_edgestate}
\end{equation}
where $\ket{G}$ denotes the many-body vacuum state, $\alpha$ and $\beta$ are constant amplitudes on sublattices $A$ and $B$, respectively, and $\mathcal{N}$ is a normalization factor. For $|\lambda|<1$ ($|\lambda|>1$), the wavefunction decays (grows) exponentially with increasing unit-cell index $n$, corresponding to localization at the left (right) boundary of the chain [see Fig.~\ref{TopologicalModels}a)].

For QST, the edge state wavefunction must be an eigenstate of the Hamiltonian in equation \eqref{eq:HamiltonianSSH}. Substituting the ansatz into the time-independent Schrödinger equation within the single-particle subspace yields conditions under which the edge states are fully supported on a single sublattice. Specifically, one finds a left-localized state supported entirely on sublattice $A$ with decay factor $\lambda_A=-J_1'/J_1$, and a right-localized state supported on sublattice $B$ with $\lambda_B=-J_1/J_1'$ \cite{Asboth2016,chen2020},
    \begin{align}
    \ket{\psi_A}&\propto\sum_{n=1}^{N/2}\lambda_A^n\hat{a}_n^\dagger\ket{G},\nonumber\\
    \ket{\psi_B}&\propto\sum_{n=1}^{N/2}\lambda_B^{N/2+1-n}\hat{b}_n^\dagger\ket{G}.
    \end{align}
In a finite chain, these two mid-gap states are not exactly degenerate, leading to a residual hybridization between the left- and right-localized modes. This hybridization energy splitting is given by
\begin{equation}
    \delta=\bra{\psi_A}\hat{H}_\mathrm{SSH}\ket{\psi_B}\propto \left(J_1'/J_1\right)^{-N/2},
    \label{eq:matrixelement}
\end{equation}
which vanishes only in the thermodynamic limit. As a result, the true eigenstates of the finite system are symmetric and antisymmetric superpositions of the two edge modes, $\ket{\psi_{\pm}}\!\!\!=\!\!\!1/\sqrt{2}\left(\ket{\psi_A}\pm\ket{\psi_B}\right)$. Consequently, for an SSH chain with an even number of sites, strictly single-sided edge eigenstates arise only in the limiting case $J_1'=0$, where the hybridization energy vanishes identically, as evident from equation \eqref{eq:matrixelement}. In this limit, the two mid-gap states become fully localized on opposite boundaries of the chain, as illustrated in Fig.~\ref{TopologicalModels}a).

\subsection{Odd SSH model, symmetry-enforced}

To relax the constraint arising from the finite-size hybridization described above, one may consider an SSH chain with an odd number of lattice sites. For a chiral symmetric Hamiltonian, the energy spectrum is symmetric about zero energy. As a consequence, systems with an odd number of sites necessarily host a zero-energy eigenstate independently of the specific values of the Hamiltonian parameters. This zero-energy mode lies within the bulk band gap in both the topologically trivial and non-trivial phases. For $J_1'\neq J_1$, the corresponding wavefunction is exponentially localized at a single boundary as seen in Figure \ref{TopologicalModels}b). Due to the sublattice imbalance introduced by the odd number of sites and the associated breaking of translational symmetry, this edge state resides entirely on sublattice $A$. 


\subsection{Rice-Mele model, controlled symmetry-breaking}

An alternative route to single-sided edge states in chains of even length is to introduce a sublattice energy offset $\hbar\Delta$, transforming the SSH Hamiltonian into the Rice-Mele model
\begin{align}
    \hat{H}_\mathrm{RM} =&\, \hbar\sum_n^{N/2} J_1'\hat{a}_n^\dagger \hat{b}_n+\hbar\!\!\sum_n^{N/2-1}\!\!J_1\hat{b}_n^\dagger \hat{a}_{n+1}\! + \textrm{h.c.}\nonumber\\&+ \hbar\Delta\sum_n^{N/2}\left( \hat{a}_n^\dagger \hat{a}_n-\hat{b}_n^{\dagger}\hat{b}_n\right)\!, \label{eq:HamiltonianRM}
\end{align}
which is commonly studied in the context of topological charge pumping \cite{Trautmann2024,Svendsen2024}. Here, the sign of $\Delta$ determines which sublattice is energetically favoured. The corresponding bulk momentum space Hamiltonian reads
\begin{equation}
         h_\mathrm{RM}(k) = \hbar\!
    \begin{pmatrix}
        \Delta && n(k) \\ 
        n^*(k) && -\Delta
    \end{pmatrix}
    \!.
\end{equation}
The introduction of the sublattice energy offset explicitly breaks chiral and particle-hole symmetries, placing the model in the AI symmetry class \cite{Ryu2010}, for which no topological invariant exists in one dimension. However, provided that $\hbar\left|\Delta\right|$ remains small compared to the bulk energy gap, edge states inherited from the topological phase of the SSH model ($|J_1'|<|J_1|$) persist. The finite sublattice energy imbalance lifts the degeneracy of the mid-gap states and shifts them to finite energies determined by $\hbar\Delta$. As a result, the edge states occupy a single sublattice and localize either on the left or right boundary of the chain, as illustrated in Figure \ref{TopologicalModels}c). 

We have now identified several mechanisms by which single-sided edge states arise in one-dimensional SSH-type models, a key requirement for robust edge-to-edge QST. While finite-size hybridization prevents strict boundary localization in even-length SSH chains, this limitation can be lifted either by considering odd-length chains, where chiral symmetry enforces a single zero-energy edge mode, or by introducing a controlled sublattice energy offset as in the Rice–Mele model. In the following sections, we show how these concepts can be implemented in Rydberg atom arrays, where long-range dipole–dipole interactions and precise geometric control enable both time-independent and time-dependent topological QST protocols.

\section{Extended SSH/Rice-Mele model with Rydberg atoms}

\begin{figure}[t]
    \centering
    \includegraphics[width =\columnwidth]{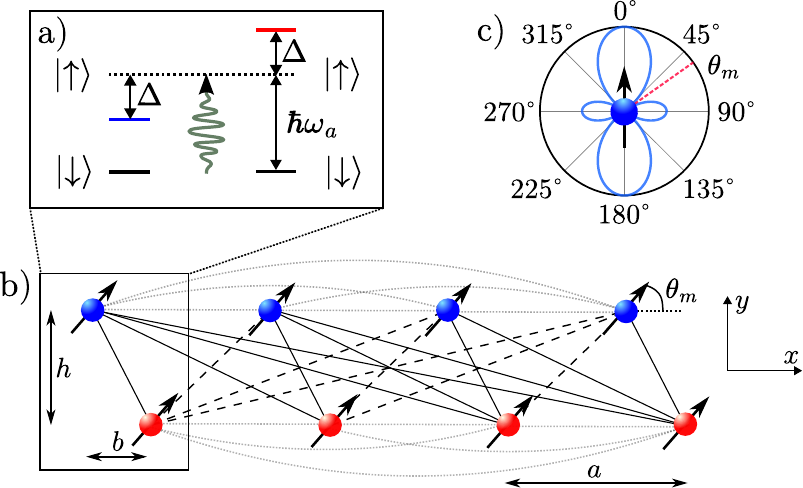}
    \caption{\textit{Realization of topological models with Rydberg atoms}. a) Each Rydberg atom is modeled as a two-level system with transition energy $\hbar\omega_a$. A staggered detuning realizes an alternating on-site potential, with atoms on sublattice $A$ (blue) and $B$ (red) experiencing a detuning $\hbar\Delta$ and $-\hbar\Delta$, respectively. b) One-dimensional chain of Rydberg atoms arranged in a bipartite lattice composed of two sublattices, $A$ and $B$. A single Rydberg excitation can hop between sites via dipole–dipole interactions, indicated by solid, dashed, and dotted lines corresponding to different coupling ranges. Sublattice $B$ is displaced relative to sublattice $A$ by a distance $b$ along the $x$-direction and $h$ along the $y$-direction. The transition dipole moments are aligned by an external field and form an angle $\theta=\theta_m$ with the $x$-axis. c) Angular dependence of the dipole–dipole interaction strength. At the angle $\theta_m=\arccos(1/\sqrt{3})$, the dipole–dipole coupling vanishes.}
    \label{RydbergTopology}
\end{figure}

In this work, we focus on the realization of topological lattice models using Rydberg atoms trapped in optical tweezer arrays. We consider $N$ Rydberg atoms, each modeled as a two-level system with both the lower ($\ket{\downarrow}$) and upper ($\ket{\uparrow}$) states belonging to the Rydberg manifold [see Fig. \ref{RydbergTopology}a)]. In typical Rydberg atom systems, the dominant decay channel of the Rydberg states involves spontaneous decay to the electronic ground state \cite{ARC1017}. In the following, we restrict our analysis to time scales well below the Rydberg states lifetimes, allowing us to neglect dissipation and treat the dynamics as purely coherent, governed by the Hamiltonian
\begin{equation}
    \hat{H}=\hbar\!\sum_{i\neq j=1}^N \!V_{ij}\hat{\sigma}_i^\dag\hat{\sigma}_j, \label{eq:HamiltonianRydberg}
\end{equation}
where $\hat{\sigma}_i=\left|\downarrow_i\right>\left<\uparrow_i\right|$ and $\hat{\sigma}^\dagger_i=\left|\uparrow_i\right>\left<\downarrow_i\right|$ are spin-1/2 ladder operators. This Hamiltonian describes the excitation exchange between atoms mediated by dipole-dipole interactions with strength
\begin{equation}
V_{ij}=\frac{d^2}{4\pi\epsilon_0\hbar} \frac{3\cos^2{\theta_{ij}}-1}{r_{ij}^3},
\label{eq:DipoleDipoleInteractionRydberg}
\end{equation}
where $\mathbf{r}_{ij}=r_{ij}\hat{\mathbf{r}}_{ij}$ denotes the interatomic separation, $\mathbf{d}=d\hat{\mathbf{d}}$ is the transition dipole moment, and $\cos{\theta_{ij}}=\hat{\mathbf{r}}_{ij}\cdot\hat{\mathbf{d}}$.

We now show how the SSH and Rice-Mele lattice structures discussed above can be realized in a Rydberg atom platform. To this end, we arrange the atoms in a one-dimensional chain composed of two sublattices, $A$ and $B$, forming a bipartite geometry analogous to the SSH model. As illustrated in Fig.~\ref{RydbergTopology}b), sublattice $B$ is displaced relative to sublattice $A$ by a distance $b$ along the $x$-direction and $h$ along the $y$-direction. An external magnetic field aligns the transition dipole moments of the Rydberg atoms, which form an angle $\theta$ with the $x$-axis, controlling both the strength and anisotropy of the dipole–dipole interactions.

Throughout this work, we restrict our analysis to the single-excitation regime, where at most one excitation $\ket{\uparrow}$ is present in the system (equivalent to a hard-core boson or spinless fermion mapping in this limit). In this limit, the spin-1/2 operators appearing in the Hamiltonian \eqref{eq:HamiltonianRydberg} can be mapped to bosonic creation and annihilation operators via a Holstein–Primakoff transformation, yielding an effective quadratic Hamiltonian of the form
    \begin{align}
        &\hat{H}_{\text{eSSH}}=\hbar\bigg\{\sum_{n=1}^{\lfloor N/2\rfloor}\sum_{m=0}^{\lfloor N/2\rfloor -n}\!\!\!J_{2m+1}'\hat{a}_n^\dagger\hat{b}_{n+m}\nonumber\\&+\!\!\sum_{n=1}^{\lfloor N/2\rfloor-1}\sum_{m=1}^{\lceil N/2\rceil -n}\!\!\left(J_{2m-1}\hat{b}_n^\dagger\hat{a}_{n+m}+J_{2m}\hat{a}_n^\dagger\hat{a}_{n+m}\right)\nonumber\\&+\!\!\sum_{n=1}^{\lfloor N/2\rfloor-1}\sum_{m=1}^{\lfloor N/2\rfloor -n}\!\!\!\!J_{2m}\hat{b}_n^\dagger\hat{b}_{n+m}+\text{h.c.}\bigg\},\label{eq:ExtendedSSH}
    \end{align}
where the coupling coefficients $J_{2m+1}'$, $J_{2m-1}$ and $J_{2m}$ are determined by the dipole-dipole interaction in equation \eqref{eq:DipoleDipoleInteractionRydberg}, describing couplings between sublattices $A$ to $B$, from $B$ to $A$, and within the same sublattice, respectively.
The resulting Hamiltonian \eqref{eq:ExtendedSSH} constitutes an extended SSH (eSSH) model that naturally incorporates long-range, in principle all-to-all, hopping processes. In the limit where only nearest neighbour couplings are retained, it reduces to the standard SSH Hamiltonian in equation \eqref{eq:HamiltonianSSH}. The presence of long-range couplings, however, leads to qualitative modifications of both the bulk spectrum and the edge state structure, which play a central role in enhancing the performance of the QST protocols that will be studied in later sections.

The bulk momentum space Hamiltonian for the eSSH model can be written as
\begin{equation}
         h_\mathrm{eSSH} (k) = 
    \begin{pmatrix}
        n_0(k) && n(k) \\ 
        n^*(k) && n_0(k)
    \end{pmatrix}
    \!,
    \label{eq:BulkHamiltonianExtendedSSH}
\end{equation}
where the off-diagonal and diagonal contributions are given by
\begin{equation*}
\begin{aligned}
    & n(k) = \hbar\!\!\sum_{m=1}^{\lfloor N/4 \rfloor}\!\!\!\left(J_{2m-1}e^{i k m a}\!+\!J'_{2m-1}e^{-i k (m-1) a}\right)\!,\\
    & n_0(k) = 2\hbar\!\!\sum_{m=1}^{\lfloor N/4 \rfloor}\!\! J_{2m}\cos(k m a).
\end{aligned}
\end{equation*}
The diagonal term  $n_0(k)$ originates from intrasublattice couplings, $J_{2m}$, and explicitly breaks chiral and particle-hole symmetries. However, this term can be suppressed by exploiting the angular dependence of the dipole-dipole interaction \eqref{eq:DipoleDipoleInteractionRydberg}. Specifically, when the transition dipole moments are aligned at the angle $\theta=\theta_m=\arccos(1/\sqrt{3})$ [see Fig. \ref{RydbergTopology}c)], the intrasublattice couplings vanish, resulting in $n_0(k)=0$. In this symmetry-preserving configuration, the eSSH Hamiltonian recovers chiral symmetry and exhibits a topological phase transition governed by the effective long-range couplings
\begin{equation}
    \begin{aligned}
       \bar{J}' &= \!\!\sum_{m=1}^{\lfloor N/4 \rfloor}\!(-1)^{m+1}J_{2m-1}', \\
       \bar{J} &= \!\!\sum_{m=1}^{\lfloor N/4 \rfloor}\!(-1)^{m+1}J_{2m-1}.
    \end{aligned}
\end{equation}
The system is in a topologically non-trivial phase for $\bar{J}'<\bar{J}$ and in a trivial phase for $\bar{J}'>\bar{J}$. Importantly, the values of $\bar{J}$ and $\bar{J}'$ can be continuously tuned by adjusting the geometric offset parameters $b$ and $h$, enabling controlled transitions between the two phases.

As in the nearest neighbour SSH model, the topologically non-trivial phase of the extended SSH model is characterized by the presence of edge states. However, the long-range couplings induce correlations between the edges and the bulk. As a result, the edge states have a finite energy that decays algebraically with the number of atoms. Similarly, the edge states may decay algebraically into the bulk rather than exponentially \cite{lepori2017}. Note that here, we will determine the localization of the edge states numerically. Chains with an odd number of sites host edge states localized on a single boundary in both the topological ($\bar{J}'<\bar{J}$) and trivial ($\bar{J}'>\bar{J}$) phases. In contrast, for chains with an even number of sites, strictly single-sided edge states occur only in the limiting case where all intercell couplings from $A$ to $B$ vanish, that is $J'_{2m-1}=0$ for all $m$.

An extended version of the Rice-Mele model with an even number of atoms can also be realized in the Rydberg array by introducing an alternating on-site potential $\hbar\Delta$, corresponding to a staggered detuning of the $\ket{\downarrow} \rightarrow \ket{\uparrow}$ transition, as illustrated in Fig. \ref{RydbergTopology}a). This results in an additional term in the Hamiltonian \eqref{eq:HamiltonianRydberg}
\begin{equation}
    \hat{H}=\hbar\!\!\sum_{i\neq j=1}^N \!\!V_{ij}\hat{\sigma}_i^\dag\hat{\sigma}_j -\hbar\Delta\sum_{j=1}^N (-1)^j\hat{\sigma}_j^\dag\hat{\sigma}_j. 
\end{equation}
Performing the same Holstein-Primakoff transformation as above, the Hamiltonian may be written as 
\begin{equation}
    \hat{H}_{\text{eRM}} = \hat{H}_\text{eSSH}+ \hbar\Delta\!\sum_{n=1}^{N/2} \!\left(\hat{a}^\dagger_n\hat{a}_n-\hat{b}^\dagger_n\hat{b}_n\right)\!.
    \label{eq:ExtendedRM}
\end{equation}
As in the nearest neighbour Rice–Mele model, the inclusion of the staggered potential breaks chiral symmetry, such that no topological invariant exists in one dimension. Nevertheless, provided that $|\Delta|$ remains small compared to the bulk energy gap, single-sided edge eigenstates persist, enabling controlled boundary localization also in chains of even length. These properties make the extended Rice–Mele model particularly suited for implementing time-dependent topological QST, as discussed in the following section.


\begin{figure*}[t]
    \centering
    \includegraphics[width =\linewidth]{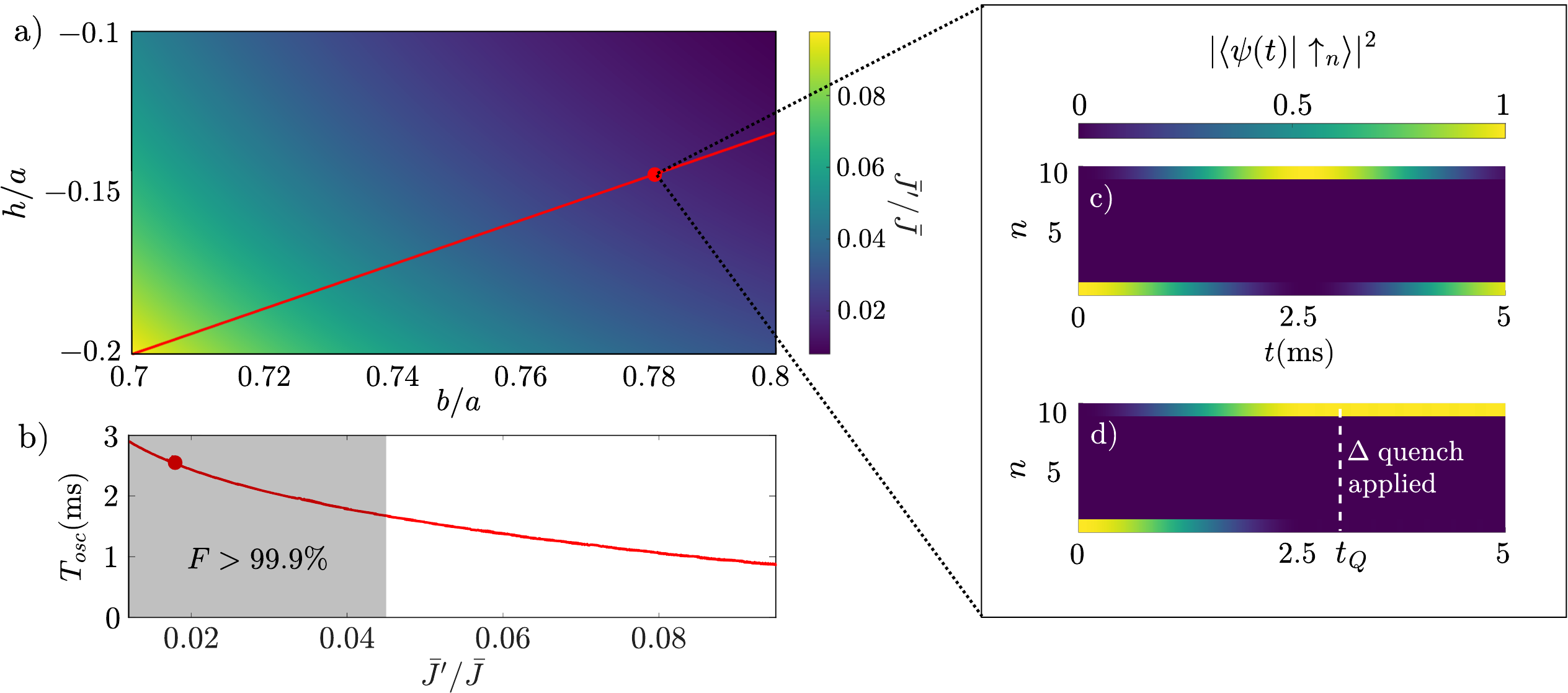}
    \caption{\textit{QST in the even extended SSH model}. a) Ratio of effective hoppings $\bar{J}'/\bar{J}$ as a function of the geometric parameters $b$ and $h$. b) Time $T_{\mathrm{osc}}$ required for a single Rabi flop between the two edges evaluated along the red line indicated in panel a). The grey shaded region marks parameter values for which the target fidelity $F(T_\mathrm{osc})>99.9\%$ is achieved. c) Dynamics of the excitation density $|\braket{\psi(t)|\uparrow_n}|^2$ of site $n$. This shows Rabi flopping dynamics for the chain configuration $b=0.78a$ and $h=-0.145a$, showing coherent oscillations between the two boundaries as a function of time $t$. d) Same dynamics as in panel c), but with a sublattice energy quench $\hbar\Delta$ applied at time $t_Q=2.635 \,\mathrm{ms}$, when the excitation reaches maximal localization on the opposite edge, suppressing further oscillations and stabilizing the transfer.}
    \label{RabiFlopping}
\end{figure*}

\section{Robust quantum state transfer in Rydberg models}

We have shown that, in the single excitation regime, a bipartite chain of Rydberg atoms with dipole-dipole couplings realizes an eSSH Hamiltonian that naturally includes long-range hopping processes. By introducing an alternating on-site potential through a staggered detuning, this system can be mapped onto an extended Rice–Mele model. In both cases, parameter regimes exist in which single-sided edge eigenstates are supported. 

In this section, we demonstrate how robust, high fidelity edge-to-edge QST can be achieved in three distinct settings: (i) the extended SSH model with an even number of sites, (ii)  the extended SSH model with an odd number of sites, and  (iii) the extended Rice-Mele model with an even number of sites. For the even-length eSSH chain, time-independent QST is realized via coherent oscillations between edge states, controlled by a quench of the sublattice energy offset. In contrast, time-dependent QST in the odd-length eSSH chain and the even-length extended Rice–Mele chain is achieved through adiabatic evolution of the Hamiltonian parameters, which connects single-sided edge eigenstates localized at opposite boundaries. In these time-dependent protocols, the transfer is initialized by preparing a single excitation on the left edge of the chain. The transfer is driven by varying the lattice parameters $b$ and $h$, corresponding to a controlled displacement of sublattice $B$ while keeping sublattice $A$ fixed. For the case of the Rice-Mele chain, this protocol is supplemented by a simultaneous variation of the staggered detuning $\Delta$. Provided the evolution is sufficiently slow, the excitation adiabatically follows the instantaneous edge eigenstate and is transported to the right boundary.

To quantitatively compare the performance of the different transfer protocols, we now specialize to a concrete and experimentally relevant Rydberg atom implementation. We consider ${}^{87}\text{Rb}$ atoms trapped in optical tweezers with a fixed lattice spacing of $a = 12\mu\text{m}$. The two Rydberg levels are chosen as the $\ket{60S_{1/2}}$ and $\ket{60P_{1/2}}$ states, for which the coupling strengths are in the order of a few MHz. These parameters are chosen for definiteness and experimental relevance. However, the transfer mechanisms discussed below are generic and remain effective for a wide range of lattice geometries, Rydberg levels, and interaction strengths, provided the system remains in the single-excitation regime and the relevant spectral gaps are preserved. Throughout the following, the dipole angle is fixed to $\theta_m = \arccos(1/\sqrt{3})$, such that $J_{2m}=0$ and chiral and particle-hole symmetries are preserved.

We characterize the efficiency of the transfer by defining a transfer time $T_{\text{transfer}}$, taken as the minimal time required to transport the initial excitation from the left boundary to the right boundary with a fidelity $F(T)=|\braket{\psi(T)|\psi_{R}}|^2>99.9\%$, where $\ket{\psi(T)}$ denotes the state of the system after a transfer time $T$ and $\ket{\psi_R}$ is the perfectly localized right edge state. 

\subsection{Even extended SSH model}

We first consider a chain consisting of an even number of Rydberg atoms, described by the eSSH Hamiltonian \eqref{eq:ExtendedSSH}. As discussed in the previous section, for an even chain length strictly single-sided edge eigenstates exist only in the limiting case where $J'_{2m-1} = 0$ for all $m$. Due to the form of the dipole-dipole interactions between Rydberg atoms, this limit cannot be realized. Consequently, in the topologically non-trivial regime the mid-gap states take the form $\ket{\psi_{\pm}}$, i.e., symmetric and antisymmetric superpositions of left- and right-localized edge states, with a finite hybridization energy $\delta$ between them, analogous to the nearest neighbour case discussed in \eqref{eq:matrixelement}.

In this regime, the dynamics of the two mid-gap states can be accurately captured by an effective two-state approximation \cite{Longhi2019_2}. Within this approximation, an excitation initialized perfectly localized on the left boundary of the chain, $\ket{\psi_L}$, evolves in time as
\begin{equation}
\begin{aligned}
    \ket{\psi(t)}&=e^{-\frac{i}{\hbar}\hat{H}_\mathrm{eSSH} t}\ket{\psi_L}\\
    &\approx\frac{1}{\sqrt{2}}\left(e^{-\frac{i}{\hbar}\delta  t}\ket{\psi_{+}}+e^{\frac{i}{\hbar}\delta t}\ket{\psi_{-}}\right)\\
    &=\cos\left(\frac{\delta}{\hbar} t\right)\ket{\psi_L}+i\sin\left(\frac{\delta}{\hbar} t\right)\ket{\psi_R}.
    \label{eq:rabioscis}
    \end{aligned}
\end{equation}
The excitation therefore undergoes coherent oscillations between the two boundaries with a frequency determined by the hybridization energy $\delta$. This Rabi flopping enables edge-to-edge QST through the free evolution of the initially prepared edge state.

This approach, however, is subject to important limitations. These can be understood by examining how the edge state localization and the transfer time depend on the ratio $\bar{J}'/\bar{J}$. Approaching the topological phase transition, i.e., increasing $\bar{J}'/\bar{J}$, enhances the hybridization between the edge states, resulting in faster oscillations. At the same time, the edge states become less localized, leading to increased coupling to bulk modes and ultimately to a breakdown of the effective two-state description. Conversely, deep in the topological regime ($\bar{J}'\ll\bar{J}$), the edge states are well localized and weakly hybridized, allowing for high-fidelity transfer, but at the expense of long oscillation periods. This trade-off between localization and speed fundamentally limits the performance of this protocol.

Figure~\ref{RabiFlopping} illustrates this trade-off. Figure~\ref{RabiFlopping}a) shows the values of the ratio $\bar{J}'/\bar{J}$ for a range of lattice offset parameters that ensure that $\bar{J}'/\bar{J}$ remains small. As shown in Figure~\ref{RabiFlopping}b), these parameter regimes allow for QST with very high fidelity, but only at the cost of long times $T_{\mathrm{osc}}$ for the first oscillation from left to right in a chain of $N=10$ atoms. Figure~\ref{RabiFlopping}c) illustrates the resulting Rabi oscillations for the parameters $b=0.78a$ and $h=-0.145a$.

\begin{figure*}[t]
    \centering
    \includegraphics[width =\linewidth]{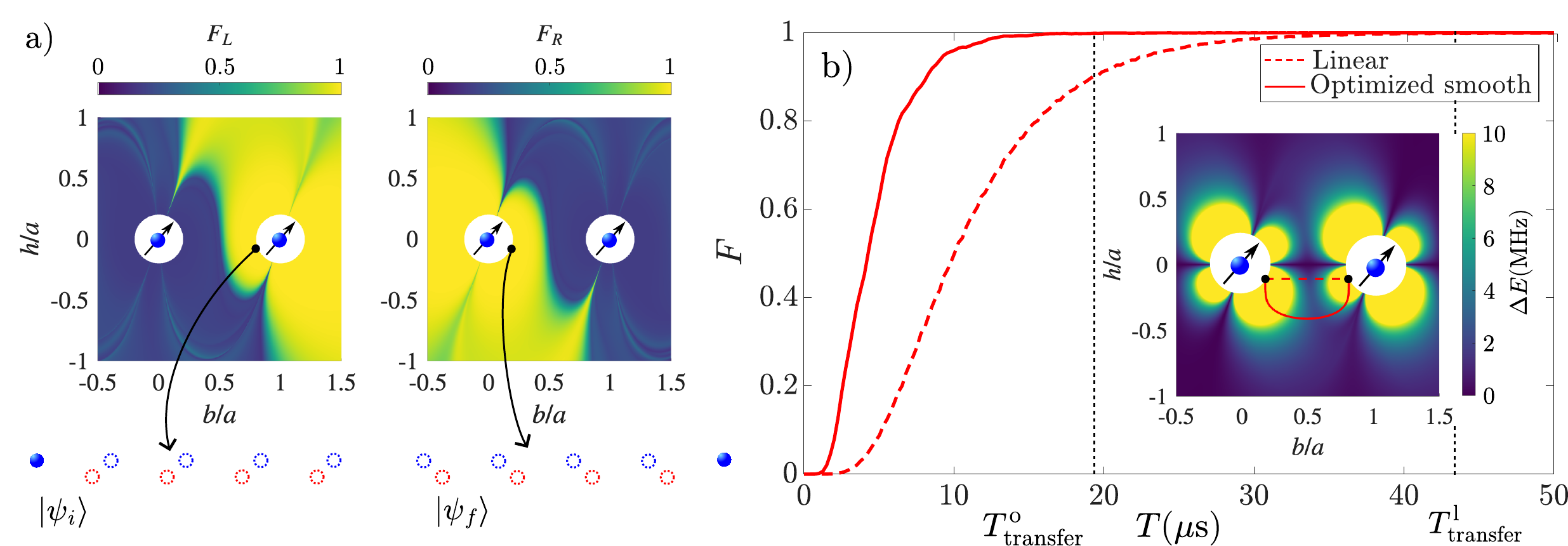}
    \caption{\textit{QST in the odd extended SSH model}. a) Overlap $F_L$ ($F_R$) between the zero-energy eigenstate $\ket{\psi_0}$ and a perfectly localized left (right) edge state as a function of the geometrical parameters $b$ and $h$. White regions indicate forbidden configurations where two tweezers are closer than $2.5\, \mathrm{\mu m}$. Black dots mark the selected initial and final chain configurations; the corresponding eigenstates $\ket{\psi_i}$ and $\ket{\psi_f}$ and associated lattice geometries for a chain of $N=9$ atoms are shown below. Dotted circles denote empty lattice sites, while filled circles indicate the presence of an excitation ($\ket{\uparrow}$). b) Transfer fidelity $F$ as a function of the total transfer time $T$ for a linear path in parameter space (dashed) and the optimized one (solid), see main text. Dotted vertical lines indicate the transfer times needed to reach the target fidelity for the optimized smooth path ($T_{\text{transfer}}^{o} = 19.5\, \mu\text{s}$) and the linear path ($T_{\text{transfer}}^{l}=43\, \mu\text{s}$). The inset shows the energy gap $\Delta E$ between the mid-gap state and the bulk states in parameter space, together with the corresponding transfer paths. The optimized path follows Eq.~\eqref{SmoothOptimizedPath} with $h_m=-0.3026a$ and $p=0.2194$.}
    \label{OddChainComb}
\end{figure*}

Moreover, since the oscillations persist indefinitely, the transfer cannot be halted without additional control. To suppress the subsequent oscillations and stabilize the transferred excitation, a non-zero sublattice energy offset $\hbar\Delta$ can be applied. As discussed in Section~3, this lifts the degeneracy of the mid-gap states and produces single-sided edge eigenstates localized at opposite boundaries. By applying a sudden quench of $\Delta$ at a time $t_Q$ when the excitation reaches maximal localization on the right boundary, the hybridization is effectively switched off and the excitation remains localized, as shown in Figure~\ref{RabiFlopping}d). While this protocol can in principle achieve near-unit transfer fidelity, the long transfer times required in the well-localized regime render it impractical for Rydberg implementations, where the achievable fidelity is ultimately constrained by the finite lifetime of the Rydberg states, here in the order of $100\,\mu\text{s}-1\,\text{ms}$.

\subsection{Odd extended SSH model}

We now turn to consider a chain consisting of an odd number of Rydberg atoms. As discussed in Section~3, chiral symmetry guarantees the existence of a zero-energy eigenstate in such a system, which for $\bar{J}'\neq\bar{J}$ is localized at a single boundary. This feature enables time-dependent QST by adiabatically connecting left- and right-localized edge states through a continuous deformation of the Hamiltonian parameters. Achieving high-fidelity transfer therefore requires identifying parameter regimes in which these edge eigenstates are strongly localized at either boundary of the chain.

In the Rydberg implementation, this localization can be controlled by varying the geometric parameters $b$ and $h$, which in turn tune the effective long-range couplings $\bar{J}$ and $\bar{J}'$. To quantify the degree of localization, we compute the overlaps
\begin{equation}
    F_{L/R}=|\braket{\psi_0 |\psi_{L/R}}|^2,
\end{equation}
where $\ket{\psi_0}$ is the zero-energy eigenstate with strongest edge character, and $\ket{\psi_{L/R}}$ denotes a perfectly localized edge state. Figure \ref{OddChainComb}a) shows $F_L$ and $F_R$ across the $(b,h)$ parameter space. The white region corresponds to geometries where two tweezers are closer than $2.5 \mu\text{m}$ and are therefore experimentally hard to achieve \cite{chew2022}. As expected, high values of $F_L$ occur when the first atom on sublattice $B$ (red) is closer to the second atom on sublattice $A$ (blue), corresponding to $\bar{J}'<\bar{J}$ and accordingly to the topologically non-trivial regime. Conversely, $F_R$ is maximized when the first atoms on sublattices $A$ and $B$ are closest, yielding $\bar{J}'>\bar{J}$ and placing the system in the trivial phase. In both cases, highly localized edge states with overlaps exceeding $99.99\%$ are obtained deep in the respective regimes, where $\bar{J}'\ll \bar{J}$ or $\bar{J}'\gg \bar{J}$. 


As a representative example, we now consider the coherent transfer of a single excitation from the leftmost to the rightmost site of an odd-length chain, as illustrated in Fig. \ref{OddChainComb}a) for a system of $N=9$ atoms. The initial chain configuration is chosen with geometric parameters $b_i = 0.812a$ and $h_i = -0.107a$, for which the mid-gap eigenstate $\ket{\psi_0}$ exhibits a strong localization on the left boundary, with an overlap $F_L>99.99\%$ with the perfectly localized left edge state. The final configuration is defined by $b_f = 0.176a$ and $h_f = h_i$, yielding a correspondingly high overlap with the perfectly localized right edge state, $F_R>99.99\%$. QST is realized by adiabatically varying $b$ and $h$ between these two configurations, such that the system remains in the instantaneous mid-gap eigenstate throughout the evolution. The achievable transfer efficiency is primarily governed by the minimum energy gap between this state and the bulk spectrum encountered along the chosen parameter path, which sets the maximum rate at which the parameters can be varied while preserving adiabaticity.

Figure \ref{OddChainComb}b) compares the transfer fidelity $F$ as a function of the total transfer time $T$ for two transfer paths in the parameter space spanned by $b$ and $h$.  The dashed line corresponds to a linear path that leaves $h$ constant while changing $b$ between the initial and final chain configurations, while the solid line represents a trajectory parameterized by
    \begin{align}
        b(t) &= b_i+\left(b_f-b_i\right)t/T,\nonumber \\
        h(t) &= h_i+h_m\sin^p{\left(\pi t/T\right)}\!,
    \label{SmoothOptimizedPath}
    \end{align}
where the parameters $h_m$ and $p$ are chosen to maximize the transfer efficiency. The inset shows the energy gap between the mid-gap state and the bulk spectrum over the relevant region of parameter space, with the dashed and solid line indicating the linear and optimized paths, respectively. The optimized trajectory is designed to follow a region of larger spectral gap throughout the evolution, thereby allowing the parameters to be varied more rapidly while maintaining adiabaticity. As a result, a target fidelity of $F>99.9\%$ is achieved after a transfer time of $T_{\text{transfer}}^{\text{o}} = 19.5\,\mu\text{s}$ for the optimized path, while the linear path requires more than twice as long, $T_{\text{transfer}}^{\text{l}} = 43\,\mu\text{s}$. 
\begin{figure}[t]
    \centering
    \includegraphics[width =\columnwidth]{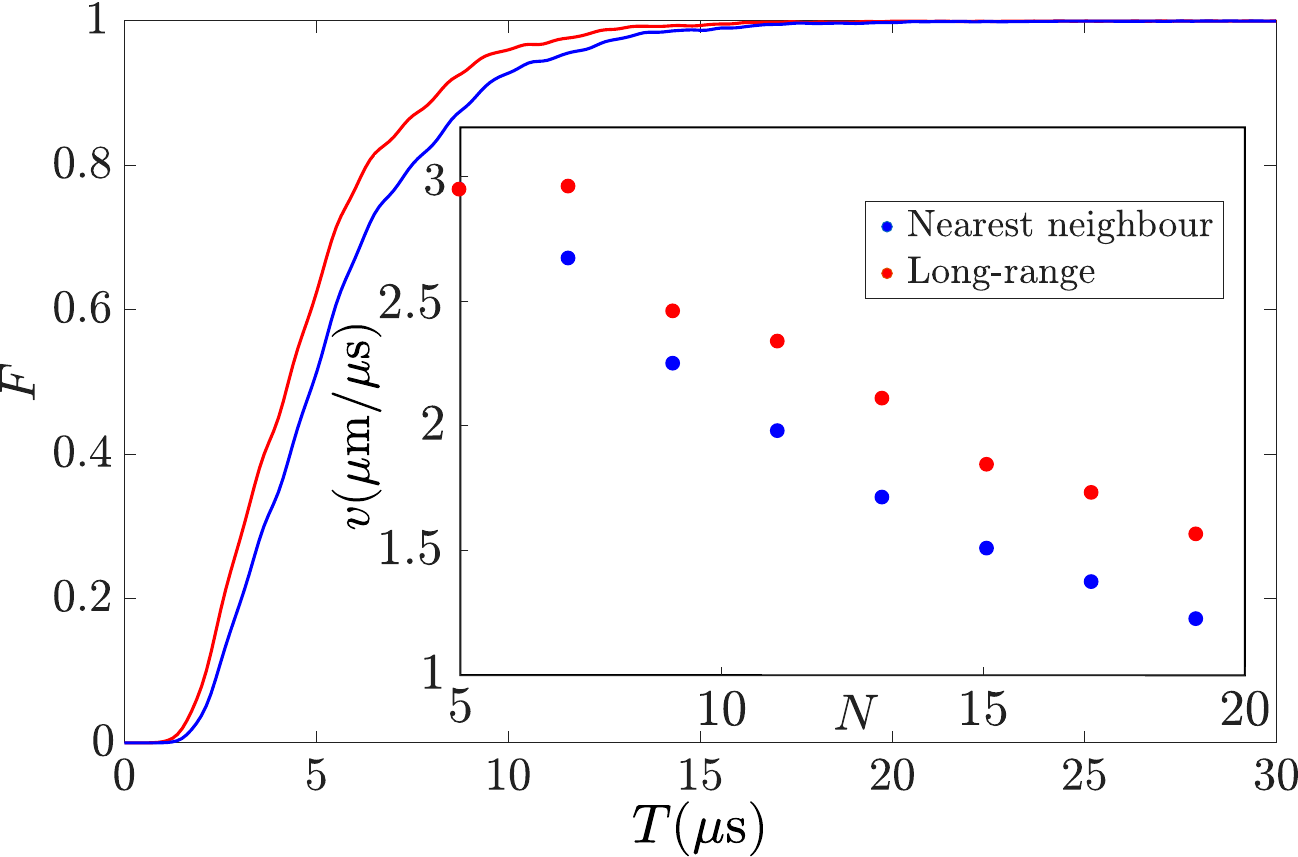}
    \caption{\textit{Long-range enhancement of transfer efficiency in the odd extended SSH model}. Transfer fidelity $F$ obtained using the optimized smooth parameter path in the odd SSH model for nearest neighbour couplings only (blue) and for the full long-range Rydberg couplings (red) for a chain of $N=9$ atoms. The parameter path was optimized indepentently for the two cases. For nearest neighbour couplings only, $h_m = -0.2957a$ and $p = 0.2194$, and for long-range couplings $h_m = -0.3026a$ and $p = 0.2194$. The inset shows the transfer velocity $v$ as a function of chain length, highlighting the enhancement induced by long-range interactions.}
\label{LRvsNNOdd}
\end{figure}

We find that the long-range character of the dipole–dipole couplings can significantly enhance the efficiency of the adiabatic transfer protocol. Figure \ref{LRvsNNOdd} compares the transfer fidelity obtained with the full long-range Rydberg interactions, described by the eSSH Hamiltonian in \eqref{eq:ExtendedSSH}, to the case where only nearest neighbour couplings are retained, corresponding to the standard SSH Hamiltonian in \eqref{eq:HamiltonianSSH}. In both cases the smooth path parametrized by \eqref{SmoothOptimizedPath} was optimized independently. The advantage of the long-range couplings becomes more pronounced when considering the transfer velocity as
\begin{equation}
    v = \frac{l}{T_\text{transfer}},
\end{equation}
with $l$ being the effective distance between left- and right-edge states, here being $l=a(N-1)/2$, which quantifies the distance traversed by the excitation per unit time. As shown in the inset of Figure \ref{LRvsNNOdd}, for a chain of $N=19$ atoms the inclusion of long-range couplings leads to a $22\%$ increase in the transfer velocity compared to the nearest neighbour case. This enhancement can be attributed to a broadening of the minimum energy gap along the adiabatic path when additional coupling terms are present, which allows for faster parameter variations while still satisfying the adiabaticity criterion.

\subsection{Extended Rice-Mele model}

\begin{figure*}[t]
    \centering
    \includegraphics[width =\linewidth]{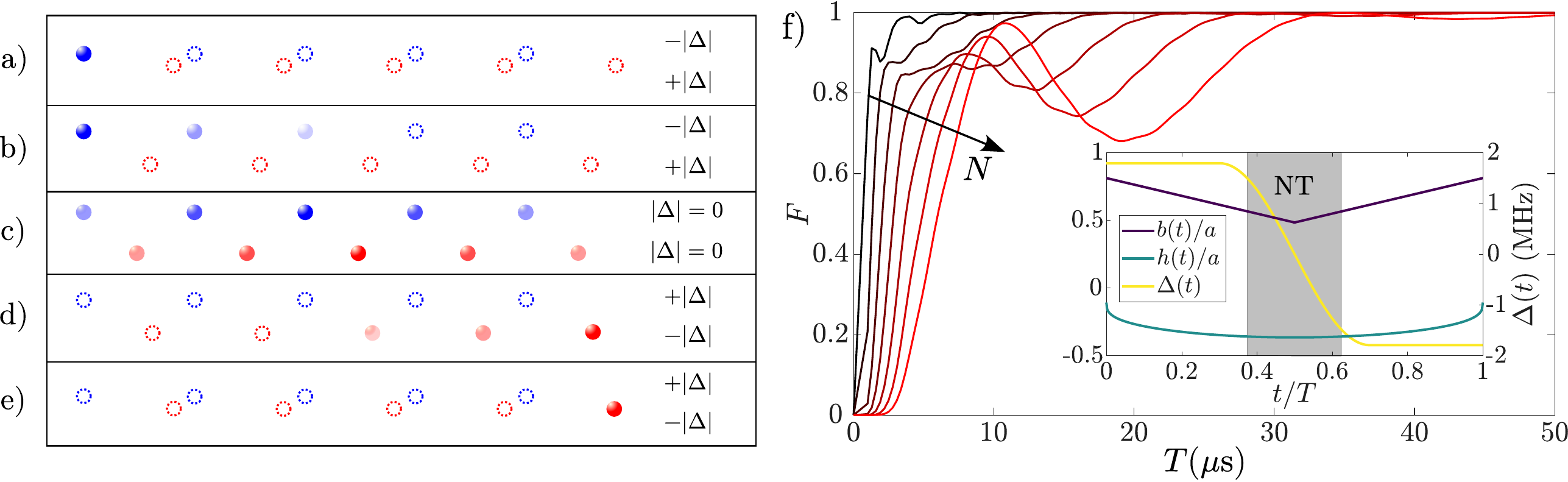}
    \caption{\textit{QST in the extended Rice-Mele model}. a)-e) Excitation probability distribution and lattice configuration at representative times during an edge-to-edge QST protocol in the extended Rice–Mele model. Empty circles denote lattice sites with zero excitation probability, while filled circles indicate non-zero excitation probability, with color intensity proportional to the local population. f) Transfer fidelity $F$ for the optimized parameter path [Eq.~\eqref{SmoothOptimizedPath} with $h_m=-0.2576a$ and $p=0.3296$] as a function of the total transfer time $T$ for increasing chain lengths from $N=4$ to $N=16$. The inset shows the temporal variation of the geometrical parameters $b$ and $h$ and of the sublattice energy offset $\hbar\Delta$ during the transfer. The grey (white) background indicates parameter regions corresponding to the non-topological (topological) phase.}
    \label{RMChainComb}
\end{figure*}
Finally, we consider time-dependent QST in a chain consisting of an even number of Rydberg atoms. As discussed above, the application of a non-zero sublattice energy offset $\hbar\Delta$ gives rise to single-sided edge states, and the system is described by the extended Rice-Mele Hamiltonian \eqref{eq:ExtendedRM}. Analogous to the odd-length eSSH chain, edge-to-edge QST in the extended Rice-Mele chain can be realized through an adiabatic variation of the geometrical parameters $b$ and $h$, supplemented by a controlled modulation of the sublattice energy shift $\hbar\Delta$.

The transfer can be understood as follows. Initially, the geometrical parameters $b$ and $h$ are chosen such that the system resides in the topological regime, $\bar{J}'<\bar{J}$. A small sublattice energy offset $\hbar\Delta$ is applied, lifting the degeneracy of the edge states without inducing coupling with the bulk states. The initial state is given by the single-sided edge eigenstate with energy $-|\hbar\Delta|$ localized on the left boundary, as illustrated in Figure \ref{RMChainComb}a). Adiabatically varying $b$ and $h$ towards the topological phase transition causes the state to delocalize over sublattice $A$ [see Figure \ref{RMChainComb}b)]. Continuing the evolution across the phase transition into the trivial regime, while simultaneously reducing $|\Delta| \rightarrow 0$, results in a state that is fully delocalized over both sublattices [Fig. \ref{RMChainComb}c)]. Reversing the sign of $\Delta$ and retracing the path in $(b,h)$ back to the initial configuration transfers the excitation to sublattice $B$, where it localizes at the opposite boundary, completing the QST [see Figs. \ref{RMChainComb}d) and e)].

The QST protocol in the extended Rice-Mele model employs the same optimized path defined in \eqref{SmoothOptimizedPath}, with the initial geometrical parameters $b_i = 0.812a$ and $h_i = -0.107a$, corresponding to the topological phase. The initial magnitude of the sublattice energy offset $|\hbar\Delta|$ is optimized to yield highly localized edge eigenstates ($F_L >99.9\%$ and $F_R >99.9\%$), and its temporal evolution during the protocol is further optimized to maximize transfer efficiency. The full protocol is illustrated in the inset of Fig.~\ref{RMChainComb}f). The resulting transfer fidelity $F$ is shown as a function of the total transfer time $T$ for increasing even chain lengths $N$. For $N>8$, the fidelity exhibits an oscillatory behavior with increasing amplitude for larger system sizes. We attribute these oscillations to non-adiabatic effects in the Rice–Mele transfer protocol \cite{Longhi2019}, where the decrease in fidelity is caused by losses to bulk states.

\begin{figure}[t]
    \centering
    \includegraphics[width =\columnwidth]{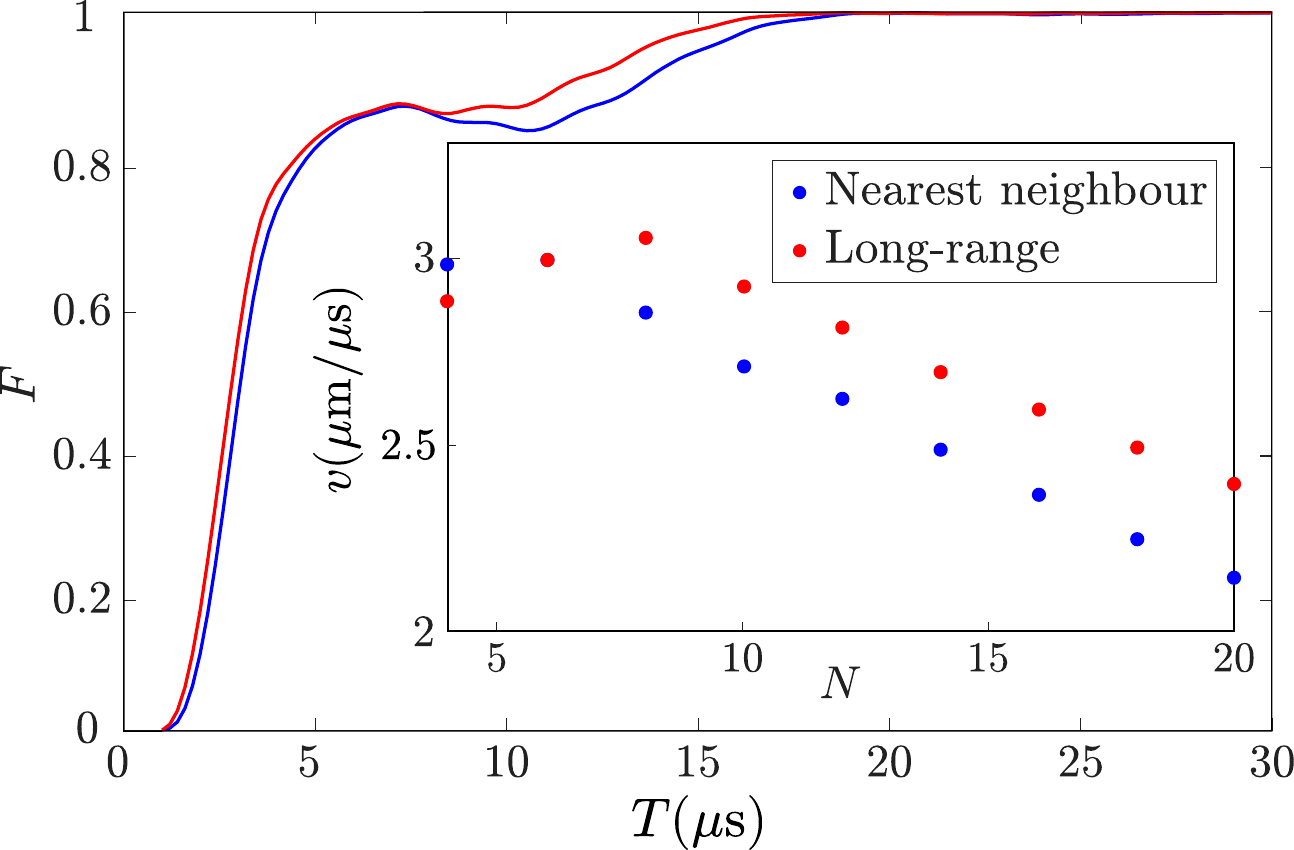}
    \caption{\textit{Long-range enhancement of transfer efficiency in the extended Rice-Mele model}. Transfer fidelity $F$ as a function of the total transfer time $T$ for the optimized smooth parameter path in the Rice–Mele model with nearest neighbour couplings only (blue) and with full long-range couplings included (red), shown for a chain of $N=10$ atoms. The parameter path was optimized for long-range couplings and nearest neighbour couplings independently and in both cases $h_m=-0.2576a$ and $p=0.3296$. The inset displays the transfer speed $v$ as a function of the chain length $N$.}
    \label{LRvsNNRM}
\end{figure}

As in the odd-length eSSH chain, the presence of long-range couplings enhances the transfer efficiency here compared to the nearest neighbour case, as shown in Figure \ref{LRvsNNRM}. The inset displays the transfer velocity $v$ as a function of the system size $N$ (here the distance $l$ is given by $l=(N/2-1)a+b$), revealing an enhancement of up to $13\%$ for $N=20$ atoms. Note that a kink appears in the transfer velocity at $N=8$ with long-range couplings, marking the crossover from an increasing to a decreasing dependence with $N$. This feature originates from the onset of oscillations in the transfer fidelity: for $N>8$, the first oscillation occurs before the fidelity reaches the threshold value $F=99.9\%$, as seen in Fig.~\ref{RMChainComb}f). In the nearest neighbour case, the corresponding kink already appears at $N=6$ atoms.

Together, these results demonstrate that Rydberg arrays with tunable geometry and detuning enable both time-independent and adiabatic topological QST protocols with high fidelity and enhanced speed, with performance benefits arising from long-range interactions. In the following section, we investigate the robustness of these schemes to local disorder, providing a realistic assessment of their performance in experimental implementations.

\section{Robustness against disorder}
\begin{figure}[t]
    \centering
    \includegraphics[width =\columnwidth]{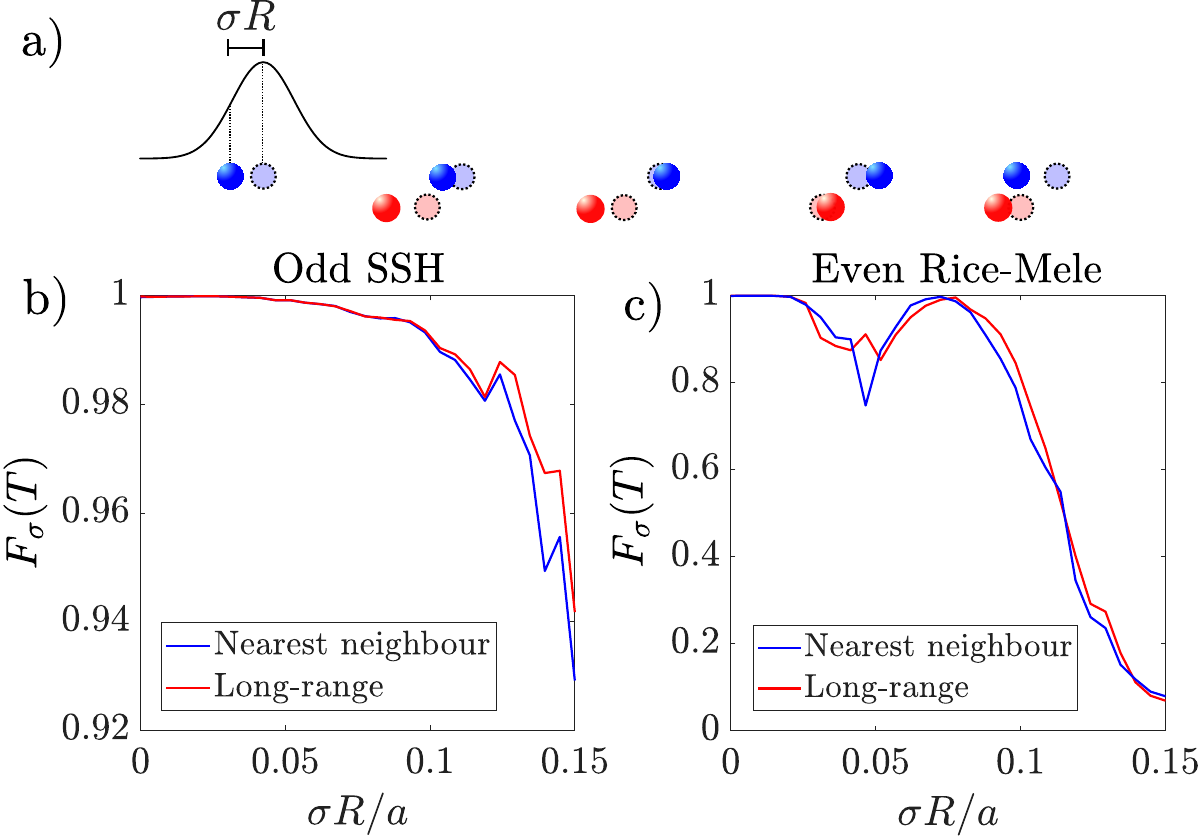}
    \caption{\textit{Robustness against positional disorder}. a) Positional disorder is implemented by randomly displacing each atom along the chain direction according to a Gaussian distribution with standard deviation $\sigma R$. Dotted circles indicate the ideal (undisordered) lattice geometry, while filled circles show the displaced atomic positions. b) Transfer fidelity $F_\sigma(T)$ after a fixed transfer time $T=30\,\mu\text{s}$ as a function of the disorder strength $\sigma R$ for the optimized smooth transfer protocol in the odd-length SSH chain with $N=9$ atoms. c) Same as in b), but for the even-length Rice–Mele chain with $N=10$ atoms. In b) and c), results are shown for nearest neighbour couplings only (blue) and including full long-range dipole–dipole interactions (red), averaged over $10^4$ disorder realizations.}
    \label{RobustnessLRvsNN}
\end{figure}

A defining feature of topological phases is their robustness against local disorder that preserves the symmetries of the underlying Hamiltonian. In the Rydberg platform considered here, a particularly relevant source of such disorder is positional disorder, arising from imperfect trapping of atoms in their optical tweezers. We model this effect by introducing random displacements of each atom along the chain direction. Specifically, the position of each atom is shifted by a random amount drawn from a Gaussian distribution with zero mean and standard deviation $\sigma R$, centered at the nominal trap position, as illustrated in Figure \ref{RobustnessLRvsNN}a). Due to the distance dependence of the dipole-dipole couplings, these random displacements translate into disorder in the effective coupling parameters $\bar{J}$ and $\bar{J}'$. To ensure that chiral and particle-hole symmetries are preserved, we restrict the disorder to the $x$-direction, such that no intrasublattice couplings $J_{2m}\neq 0$ are induced by the disorder.

Figure~\ref{RobustnessLRvsNN} shows the resulting transfer fidelity $F_\sigma(T) = |\braket{\psi_\sigma(T) |\psi_{R}}|^2$ evaluated at a fixed total transfer time $T=30\,\mu\text{s}$, chosen such that in the no-disorder case $F_{\sigma=0}(T)>99.9\%$. Here, $\ket{\psi_\sigma(T)}$ denotes the final state averaged over multiple realizations of the positional disorder. For the odd-length SSH chain [Fig.~\ref{RobustnessLRvsNN}a)], we observe a pronounced robustness against positional disorder: transfer fidelities exceeding $93\%$ are maintained even for disorder strengths as large as $\sigma R=0.15 a$. This robustness is a direct consequence of topological protection. Remarkably, and somewhat counter-intuitively, the inclusion of long-range couplings further enhances the robustness, despite the fact that positional disorder introduces fluctuations in a larger number of coupling terms. This suggests that the gap enhancement induced by the long-range interactions outweighs the increased sensitivity to positional fluctuations. In contrast, the Rice-Mele chain is significantly more sensitive to strong disorder. For disorder strengths $\sigma R > 0.1a$, the transfer fidelity rapidly degrades. This behaviour can be understood from the structure of the transfer protocol: a successful QST requires the system to remain in the topologically trivial regime for a finite duration, during which the excitation delocalizes across the lattice. In this regime, topological protection is absent, and strong disorder leads to mixing between eigenstates, suppressing coherent transport. Nevertheless, for disorder strengths typical of current experimental conditions \cite{Browaeys2020}, the Rice-Mele protocol remains robust and yields high transfer fidelities. Note that, even when disorder in the $y$-direction is included, breaking chiral and particle-hole symmetry, our results remain qualitatively unchanged for weak disorder strengths. 

\section{Conclusion and outlook}
In this work, we have demonstrated both time-independent and time-dependent protocols for high-fidelity edge-to-edge quantum state transfer in one-dimensional chains of Rydberg atoms with  long-range dipole-dipole couplings. We have shown that the efficiency of the transfer is governed by the geometry of the  atomic chain, which effectively controls the hopping amplitudes and the size of the spectral gap. By optimizing these geometric parameters, time-dependent transfer protocols achieve fidelities exceeding $99.9\%$ for chains with both odd and even numbers of atoms, with transfer times well below the intrinsic lifetimes of the Rydberg states. Importantly, we have shown that the transfer remains robust against symmetry-preserving positional disorder within experimentally realistic tolerances. Contrary to naive expectations, the presence of long-range couplings, which are naturally present in Rydberg systems, not only enhances the transfer efficiency relative to nearest neighbour models, but can also improve the robustness of the protocol. These results highlight long-range interactions as a valuable resource for topologically assisted QST.

While the transfer efficiency in this work was significantly enhanced by optimizing the path through parameter space, further improvements are possible by additionally tailoring the temporal variation of the Hamiltonian parameters, for instance by adapting the protocol to the instantaneous size of the energy gap \cite{Palaiodimopoulos2021}. Beyond one-dimension, the concepts developed here naturally extend to higher-dimensional systems. In particular, two-dimensional Rydberg atom arrays, where topological edge states have been predicted  \cite{Weber2022,Weber2018,Svendsen2025}, provide a promising platform for realizing topologically assisted state transfer along extended boundaries.

Although dipole–dipole interactions in Rydberg systems decay cubically with the distance, incorporating these realistic long-range couplings enhances both the efficiency and robustness of QST compared to nearest neighbour models. While the conventional bulk–boundary correspondence is expected to remain valid under the conditions considered here \cite{jones2023,chen2020}, sufficiently long-range interactions are known to weaken this correspondence \cite{lepori2017,mcdonnell2022}. A natural next step is therefore to systematically investigate how the range of interactions influences topological state transfer. This could be explored in other quantum-optical platforms, such as atomic gases in low-lying electronic states or atoms coupled to nanophotonic waveguides, where interactions can become effectively infinite-ranged \cite{Defenu2023}.

Finally, an intriguing direction is the transfer of entangled states. By introducing an additional sublattice with the same vertical offset as sublattice $B$, a quantum-optical realization of an SSH3 chain \cite{xie2019} becomes feasible. Such a geometry could enable the coherent transport of multipartite or entangled states across the system \cite{Wang2022}, opening new avenues for topological quantum information processing.

\begin{acknowledgments}
We acknowledge funding from the research unit FOR5413 (Grant No. 465199066). This work is also supported by the European Research Council Grant OPEN-2QS (Grant No. 101164443). We acknowledge support by Open Access Publishing Fund of the University of Tübingen. Finally, we acknowledge support by the state
of Baden-Württemberg through bwHPC and the
German Research Foundation (DFG) through
grant no INST 40/575-1 FUGG (JUSTUS 2 cluster).
\end{acknowledgments}

\section*{AI usage} The authors used \emph{Claude (Sonnet 5)} for grammar and spell-checking of the manuscript. All suggestions were reviewed and approved by the authors.

\end{document}